\documentclass{appolb}
\usepackage{amsmath}
\usepackage{amssymb}
\usepackage{graphicx}
\usepackage[utf8]{inputenc}
\usepackage{verbatim}
\usepackage[T1]{fontenc}
\usepackage[english]{babel}

\usepackage{subfigure}

\begin{document}
\title{Momentum broadening of heavy quarks and jets in the Glasma from classical colored particle simulations
\thanks{Presented at Quark Matter 2022, Krak\'ow, Poland 
}
}
\author{Dana Avramescu\thanks{Presenter and corresponding author: {dana.avramescu@s.unibuc.ro}}, Virgil Băran \vspace{-0.8em}
\address{Faculty of Physics, University of Bucharest, 077125 Măgurele, Romania}
\\[3mm]
{Andreas Ipp, David I.~M\"uller  \vspace{-0.8em}
\address{Institute for Theoretical Physics, TU Wien, 1040 Vienna, Austria}
}
\\[3mm]
Vincenzo Greco  \vspace{-0.8em}
\address{Department of Physics and Astronomy, University of Catania, Via S.~Sofia 64, I-95125 Catania} \vspace{-0.8em}
\address{INFN-Laboratori Nazionali del Sud, Via S.~Sofia 62, I-95123 Catania, Italy}
\\[3mm]
Marco Ruggieri  \vspace{-0.8em}
\address{School of Nuclear Science and Technology, Lanzhou University,
222 South Tianshui Road, Lanzhou 730000, China}
}
\maketitle
\begin{abstract}
We investigate the effects of the pre-equilibrium Glasma stage of heavy-ion collisions on the broadening of momentum for heavy quarks and jets using classical colored particle simulations based on Wong's equations. Confirming previous studies, we find that these probes accumulate momentum on the order of the saturation scale $Q_s$ and that the color field of the Glasma induces an anisotropy with more broadening along the beam axis. For quark jets the anisotropy is more pronounced at lower energies.
\end{abstract}
  
\section{Theoretical background}

Heavy-ion collision experiments performed at the Large Hadron Collider (LHC) and the Relativistic Heavy-Ion Collider (RHIC) allow us to study QCD under extreme conditions. In particular, hard probes such as jets and heavy quarks might reveal details about the earliest stages of the evolving medium produced in nuclear collisions \cite{Andres_2020}. At very early times the medium is dominated by strong classical color fields known as the Glasma \cite{Gelis_2010}.

According to the Color Glass Condensate (CGC) effective theory \cite{Gelis_2010}, the hard partons of high energy nuclei effectively behave as Lorentz-contracted lightlike classical color currents $\mathcal J^\mu$. These currents act as sources for a highly occupied and thus classical gluon field $\mathcal A^\mu$ which describes the soft gluonic content of the nuclei. In a nuclear collision the non-Abelian interaction among the two color fields produces the Glasma, which, at leading order in the strong coupling constant $g$, is a classical color field whose dynamics is determined by the Yang-Mills (YM) equations
\begin{align}
    \mathcal D_\mu \mathcal{F}^{\mu\nu}(x^+, x^-, \mathbf x_\perp) = \mathcal{J}^\nu_A(x^-, \mathbf x_\perp) + \mathcal{J}^\nu_B(x^+, \mathbf x_\perp).
\end{align} 
Here, $A$ and $B$ denote the two nuclei, $\mathcal D_\mu$ is the gauge-covariant derivative and $\mathcal{F}_{\mu\nu}$
is the field strength tensor. In the ultrarelativistic limit, the color field of the Glasma can be chosen to be independent of rapidity $\eta = \ln(x^+ / x^-)/2$, reducing the system to effectively 2+1 dimensions. The dynamics is then governed by the source-free YM equations in the forward light cone parametrized by proper time $\tau = \sqrt{2 x^+ x^-} > 0$
\begin{equation}
    \label{ymfieldeqs}
    \partial_\tau \big(\tau\partial_\tau \mathcal A_i \big)=\tau\mathcal{D}_jF_{ji}-\dfrac{\mathrm{i}g}{\tau}\Big[\mathcal A_\eta,\mathcal{D}_i \mathcal A_\eta\Big], \quad 
    \partial_\tau \bigg( \frac{1}{\tau}\partial_\tau \mathcal  A_\eta\bigg)=\dfrac{1}{\tau}\mathcal{D}_i\big(\mathcal{D}_i \mathcal A_\eta \big),
\end{equation}
with initial conditions for $\mathcal A^\mu$ provided at the $\tau \rightarrow 0^+$ boundary.
In this work, we focus the on McLerran-Venugopalan initial conditions, which are used to model large nuclei with infinite extent in the transverse plane.

Roughly at the same time as the Glasma is created from collisions among the soft partons of the nuclei, heavy quarks such as charm and beauty and highly energetic quark and gluon jets may form from hard scatterings. These colored probes then propagate through and interact with the Glasma, which may leave characteristic imprints of the initial stages of the evolving medium at late times \cite{Ipp_2020, Khowal_2021, Carrington_2022}. Approximating these probes as classical color charges, we model their dynamics using Wong's equations 
\begin{equation}
    \label{wongcurv}
    \begin{aligned}
        &\frac{\mathrm{D}x^\mu}{\mathrm{d}\boldsymbol{\tau}}=\frac{p^\mu}{m}, \qquad
        &\frac{\mathrm{D}p^\mu}{\mathrm{d}\boldsymbol{\tau}}=gQ^a \mathcal F^{\mu\nu,a}\frac{p_\nu}{m}, \qquad
        &\frac{\mathrm{d}Q^a}{\mathrm{d}\boldsymbol{\tau}}=-gf^{abc} \mathcal A_\mu^bQ^c\frac{p^\mu}{m},
    \end{aligned}
\end{equation}
where $m$ is the particle mass, $Q^a(\boldsymbol{\tau})$ is the color charge and $ \boldsymbol{\tau}$ is the proper time in the rest frame of the particle. $D/d \boldsymbol{\tau}$ denotes the covariant derivative along the trajectory associated with the $(\tau, \eta)$ coordinate frame.
Neglecting any backreaction from the probe to the Glasma, the solution to Eq.~\eqref{ymfieldeqs} provides a color background field in which the particles evolve as ``test particles''.

We solve Eqs.~\eqref{ymfieldeqs} and \eqref{wongcurv} numerically using the colored particle-in-cell (CPIC) approach.
For each collision event we simulate an ensemble of independent particles with random color charges $Q^a$ satisfying%
\begin{align}
    \label{qnpointfct}
    \langle Q^a \rangle_R =0, \quad
    \langle Q^a Q^b \rangle_R =T_R \delta^{ab}, \quad
    \langle Q^a Q^b Q^c \rangle_R =\frac{B_R}{4}d^{abc},
\end{align}
with $T_F = 1/2$, $T_A = N_c$, and $B_F = 1$, $B_A = 0$ for fundamental ($F$, quarks) and adjoint ($A$, gluons) particles. Our main  observable of interest is the broadening of momentum
\begin{align}
    \label{mombroad}
    \delta p_\mu^2(\tau)\equiv p_\mu^2(\tau)-p_\mu^2(\tau_\mathrm{form}), \qquad \mu \in \{ x,y,z \},
\end{align}
where $\tau_\mathrm{form}$ is the formation time of the particle. 
We relate the classical momentum broadening to broadening computed within pQCD in the eikonal approximation defined via a Wilson loop \cite{DEramo_2011, Ipp_2020}
\begin{align}  
    \label{match}
    \frac{1}{D_R}\langle \langle\delta p_\mu^2(\tau) \rangle \rangle_R^\mathrm{classical} = \big\langle\delta p_\mu^2(\tau)\big\rangle_R^\mathrm{pQCD},
\end{align}
where $D_F = N_c$, $D_A = N^2_c - 1$. In the pQCD case, averaging is performed over the background field, while in the classical case we additionally average over color charges  according to Eq.~\eqref{qnpointfct}. 

\section{Results and conclusion}
\begin{figure}[tb]
\centerline{
\includegraphics[width=\textwidth]{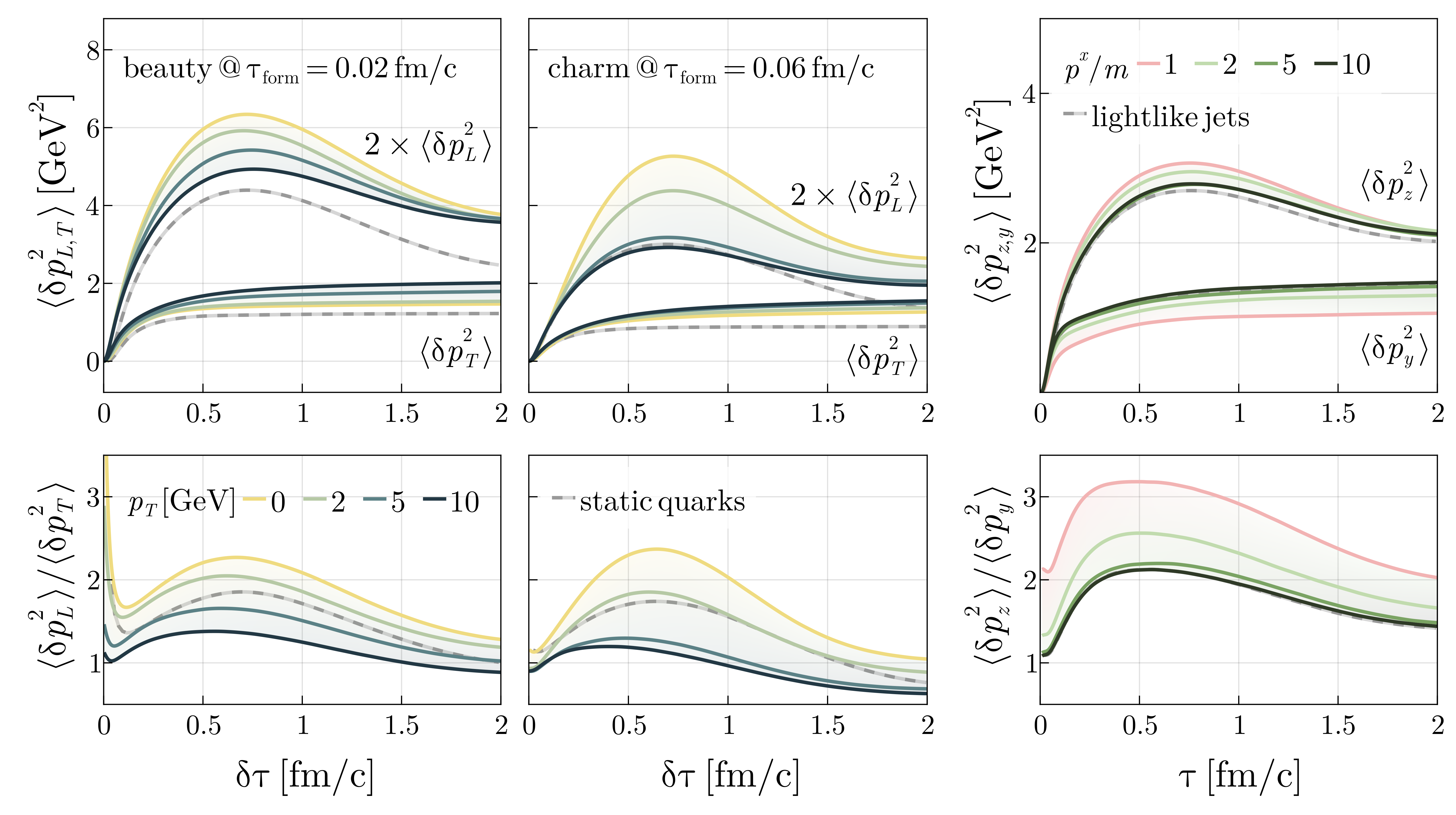}}
\caption{Left and middle: longitudinal and transverse momentum broadening of beauty and charm quarks and their ratio (anisotropy) as a function of $\delta \tau = \tau - \tau_\mathrm{form}$. Right: $y$- and $z$-broadening for jets propagating along the $x$-axis and their ratio.}
\label{Fig:wong_vs_kappa}
\end{figure}

Our results for heavy quarks and jets in the Glasma are summarized in Fig.~\ref{Fig:wong_vs_kappa}. Here, the Glasma is characterized by a saturation momentum $Q_s=2\,\mathrm{GeV}$. In the case of heavy quarks, we consider charm and beauty with $m_c = 1.27 \,\mathrm{GeV}$ and $m_b = 4.18\,\mathrm{GeV}$ formed at $\tau_{c,b} = 1 / (2 m_{c,b})$ with initial momentum in the transverse plane $p_T\in[0,10]\,\mathrm{GeV}$. Quark jets are initialized with momenta along the $x$-axis with $p^x/m\in[1,10]$ at $\tau_\mathrm{form} \rightarrow 0^+$. We also consider the limits $m \rightarrow \infty$ (static quarks) and $p^x/m \rightarrow \infty$ (lightlike quark jets), which can be computed from the temporal and lightlike Wilson loop respectively. In these limiting cases, our classical particle simulation reproduces previous results for momentum broadening in the Glasma \cite{Ipp_2020}.  We find that heavy quarks and jets accumulate squared momenta of around $\sim 4 \, \mathrm{GeV}^2$ within relatively short times of $\tau \lesssim 0.5 \, \mathrm{fm}/c$, which is comparable to $Q^2_s$. By comparing to the limiting cases we see that for dynamic heavy quarks $\langle \delta p_L^2 \rangle = \langle \delta p_z^2 \rangle$ can increase by up to $50\%$ and $\langle \delta p_T^2 \rangle  = \langle \delta p_x^2 + \delta p_y^2 \rangle$ by up to $30\%$. The anisotropy $\langle \delta p_L^2 \rangle / \langle \delta p_T^2 \rangle$ is less affected by finite mass and momentum. In the case of jets, $\langle \delta p_z^2 \rangle$ and $\langle \delta p_y^2 \rangle$ differ from the lightlike scenario by at most $30\%$, whereas the corresponding anisotropy may increase by up to $60\%$ for jets with small $p^x/m$ ratio. As demonstrated in \cite{Ipp_2020}, the momentum broadening anisotropy for lightlike jets can be traced back to characteristic correlations among the color fields of the Glasma. We have shown that this anisotropy persists if the full classical dynamics of the probes is taken into account and that such an anisotropy also exists for heavy quarks, albeit less pronounced.
In future works, we plan to compute additional observables which quantify the effect of the Glasma on colored probes such as angular correlations of quark anti-quark pairs and gauge invariant force correlators along the trajectories of the probes. 

\bibliographystyle{IEEEtran}
\bibliography{references}
\end{document}